\documentclass[11pt, conference]{IEEEtran}
\usepackage{mathptmx} 
\usepackage{titlesec}
\usepackage{parskip}
\usepackage{graphicx}
\graphicspath{ {./} }
\usepackage{cite}
\usepackage{amsmath,amssymb,amsfonts}
\usepackage{algorithmic}
\usepackage{graphicx}
\usepackage{textcomp}
\usepackage{xcolor}

\usepackage{fancyhdr}
\usepackage[normalem]{ulem}
\usepackage[hyphens]{url}
\usepackage[final]{microtype}
\usepackage[keeplastbox]{flushend}
\usepackage[bookmarks=true,breaklinks=true,letterpaper=true,colorlinks,linkcolor=black,citecolor=blue,urlcolor=black]{hyperref}

\title{Reuse Cache for Heterogeneous CPU-GPU Systems} 

\author{
\IEEEauthorblockN{Tejas Shah}
\IEEEauthorblockA{tyshah2@wisc.edu}
\and
\IEEEauthorblockN{Bobbi Yogatama}
\IEEEauthorblockA{bwyogatama@cs.wisc.edu}
\IEEEauthorblockA{}
\and
\IEEEauthorblockN{Kyle Roarty}
\IEEEauthorblockA{kroarty@wisc.edu}
\and
\IEEEauthorblockN{Rami Dahman}
\IEEEauthorblockA{rami@cs.wisc.edu}
}

\begin{document}
\maketitle
\pagestyle{plain}

\begin{abstract}
It is generally observed that the fraction of live lines in shared last-level caches (SLLC) is very small for chip multiprocessors (CMPs). This can be tackled using promotion-based replacement policies like re-reference interval prediction (RRIP) instead of LRU \cite{RRIP}, dead-block predictors \cite{dead}, or reuse-based cache allocation schemes \cite{reuse}. In GPU systems, similar LLC issues are alleviated using various cache bypassing techniques. These issues are worsened in heterogeneous CPU-GPU systems because the two processors have different data access patterns and frequencies. GPUs generally work on streaming data, but have many more threads accessing memory as compared to CPUs. As such, most traditional cache replacement and allocation policies prove ineffective due to the higher number of cache accesses in GPU applications, resulting in higher allocation for GPU cache lines, despite their minimal reuse.
	
In this work, we implement “The Reuse Cache” approach \cite{reuse} for heterogeneous CPU-GPU systems. The reuse cache is a decoupled tag/data SLLC which is designed to only store the data that is being accessed more than once. This design is based on the observation that most of the cache lines in the LLC are stored but do not get reused before being replaced. We find that the reuse cache achieves within 0.8\% of the IPC gains of a statically partitioned LLC, while decreasing the area cost of the LLC by an average of 40\%.

\end{abstract}

\section{Introduction}
    Due to the end of Dennard Scaling, heterogeneous chip multiprocessors (CMPs) are becoming increasingly more popular as a way to increase energy efficiency while avoiding performance losses to dark silicon. By having multiple types of cores on the same chip, work can be offloaded to cores that are specialized for a task, which has been shown to increase performance over typical, homogeneous CMPs \cite{commie}.
    
    In a heterogeneous CPU-GPU CMP, the CPU and GPU share both an address space and a last-level cache (LLC). This can lead to issues of cache thrashing and dead blocks in the LLC. Typically, CPU applications have significant amounts of data reuse whereas GPU applications have massively parallel thread blocks that access significant amounts of memory with minimal reuse. This indicates that GPUs often end up overloading the LLC, evicting useful cache lines that the CPU would reuse in the future. This cache thrashing can significantly decrease the performance of the CPU in the system \cite{commie}.
    
    The reuse cache is a decoupled tag and data cache that only allocates cache blocks that are being referenced more than once. Prior works have implemented the reuse cache design in the last level cache of multi-core CPU \cite{reuse}. We propose to adopt the reuse cache in a CPU-GPU system. We believe that due to the streaming nature of GPU applications, adopting the reuse cache could limit the amount of the GPU accesses that enter the LLC and reduce the number of dead blocks while significantly decreasing the LLC power and area cost.
    
    Our key contributions are as follows:
    \begin{itemize}
        \item We apply the reuse cache approach to heterogeneous CPU-GPU systems. This approach has been previously applied to homogeneous CMPs, but not for heterogeneous systems.
        \item We analyze the performance and area impact of this system compared to two other popular caching schemes for shared LLCs in heterogeneous systems.
    \end{itemize}



\section{Related Work}

    Several LLC caching schemes exist for heterogeneous CPU-GPU systems. In this section, we present each and analyze their benefits and drawbacks.

    \subsection{Cache Partitioning}
        In order to protect the CPU's cache lines from being evicted by the GPU, many have proposed partitioning the LLC into two sections, one for the CPU and one for the GPU \cite{static}. This way, the GPU will only evict the block from its partition, leaving the CPU's partition unaffected. Therefore, this approach could completely eliminate the cross-core thrashing between CPU and GPU. This can be done statically by pre-defining the cache partitions, or dynamically by taking into account program and hardware information \cite{UCP} \cite{TAP}. However, static partitioning can limit the effective LLC size, and dynamic partitioning requires additional hardware in order to both predict cache usage and adjust cache partitions.
        
    
    
    \subsection{Cache Bypassing}
        Since the nature of GPU applications is mostly streaming without reuse, prior work has proposed bypassing the LLC entirely for GPUs \cite{bypass}. Even though this approach could remove all cross-core cache thrashing in the LLC, GPU applications that are not streaming-based and do have data reuse will not benefit from this approach. 
        

\begin{figure}
    \centering
    \includegraphics[width=0.8\columnwidth]{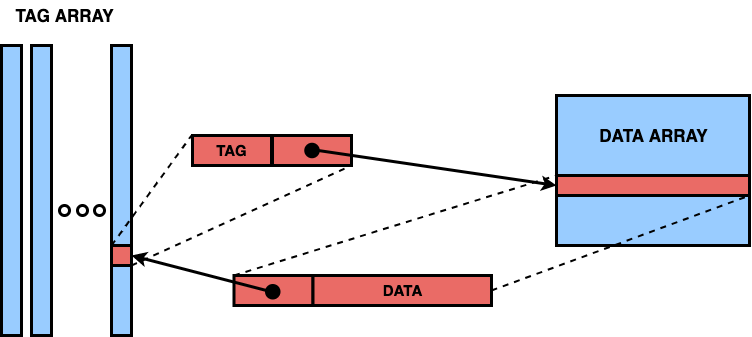}
    \caption{Tag entry and data entry in the reuse cache}
    \label{figs:pointer}
\end{figure}

\begin{figure}
    \centering
    \includegraphics[width=\columnwidth]{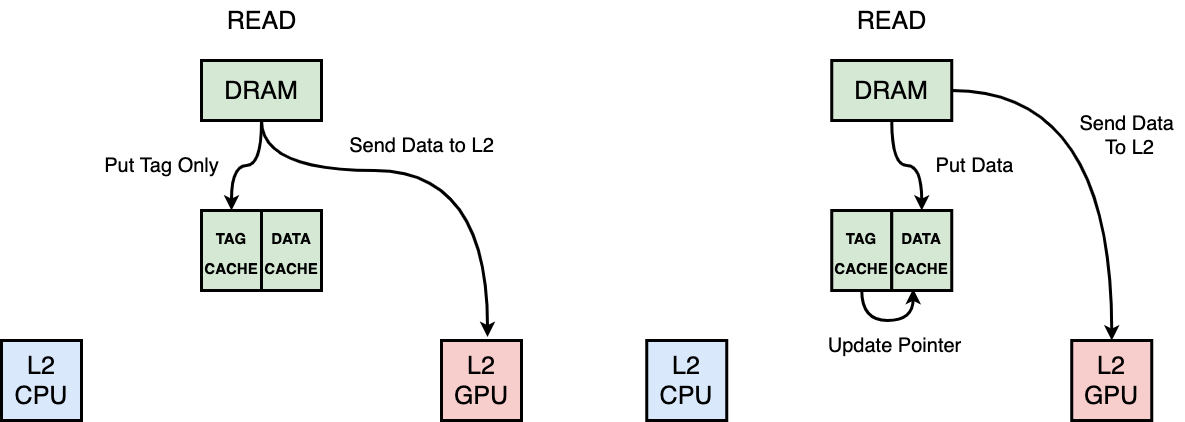}
    \caption{(a) The first access in the reuse cache (tag insertion) (b) The second access in the reuse cache (data insertion)}
    \label{figs:reuse}
\end{figure}


\section{The Reuse Cache Design}
    The reuse cache is designed to prevent dead blocks in the LLC by only allowing data to be placed in the cache if it is accessed more than once. In order to keep track of which lines have been re-referenced, the LLC's tag array is decoupled from the data array. Each tag array entry contains a tag as well as a pointer to the corresponding data in the data array. The data array would also contain a reverse pointer to the tag array entry to set the tag entry pointer to NULL during data eviction. Figure \ref{figs:pointer} shows the tag array and the data array entry of the reuse cache.
    
    Figure \ref{figs:reuse} shows an example of the first access and second access in the reuse cache. During the first LLC miss (Figure \ref{figs:reuse}(a)), after getting the data from the memory, the tag is then looked up in the tag array. Since the tag is not found, it is added to the tag array and the pointer is set to NULL. The fetched data is then sent directly to the private cache without being placed in the LLC. Figure \ref{figs:reuse}(b) shows a second LLC miss on the same address. In this case, the tag is already present in the tag array and the address is considered as being reused. Upon noticing the tag hit, the data that is brought from the memory will then be placed into the data array and the tag pointer will be updated accordingly. Subsequent accesses after this step will be a cache hit.
    
    For LLC eviction, the tag and data arrays are handled separately. During tag eviction, the tag pointer is checked to determine if the corresponding data is present in the data array. If it does, the cache line will also be evicted from the data array. Otherwise, only the tag array's line is evicted. On the other hand, during data eviction, the reverse pointer in the data array will look for the corresponding tag entry. The tag pointer will then be set to NULL and the data entry will be evicted from the LLC.
    

\section{Experimental Methodology}
We used the AMD APU model in the gem5\cite{gem5} simulator for implementing static partitioning (50:50 cache ratio between CPU-GPU), GPU bypassing, and reuse-cache policies. The system parameters used in gem5 were as follows:
\begin{scriptsize}
\begin{table}[h!]
  \centering
  \begin{tabular}{|l|l|}
    \hline
    \textbf{CPU} &  \\
    \hline
    Core & Dual-core TimingSimple CPU \\
    \hline
    L1 Cache & Private 4-way, 64B line, 32KB I/D,\\ 
     & PseudoLRU \\
    \hline
    L2 Cache & Unified 8-way, 64B line, 256KB,\\  & PseudoLRU\\
    \hline
    \hline
    \textbf{GPU} &  \\
    \hline
    Core & 4 compute units (CUs), \\
     & 64-thread wavefront \\
    \hline
    L1 Cache & 64B line, PseudoLRU, Writethrough\\ 
     & Private 4-way, 4KB Data, \\ 
     & Shared 8-way, 32KB Instruction \\
     & (4 CUs per instruction cache)\\
    \hline
    L2 Cache & Unified 8-way, 64B line, 4KB, \\  & PseudoLRU, Writeback \\
    \hline
    \hline
    \textbf{Shared Components}\\
    \hline
    LLC & 16-way, 64B line, 512KB-4MB, LRU\\
    \hline
    DRAM & 2GB\\
    \hline
  \end{tabular}
  \newline
  \caption{Configuration for the heterogeneous system evaluation}
\end{table}
\end{scriptsize}

Heterogeneous workloads were prepared using different combinations of CPU/GPU benchmarks shown in table \ref{benchmarks}, taking into account their memory access behaviour. The benchmark classifications are determined using a parameter called "reuse-distance", which quantifies how frequently an address is accessed by the program.

\begin{scriptsize}
\begin{table}[h!]
  \centering
  \begin{tabular}{|l|l|}
    \hline
    \textbf{CPU Benchmarks} &  \\
    \hline
    Cache-friendly & Queens, SHA \\
    \hline
    Cache-sensitive & Blocked matrix-multiplication \\
    \hline
    Large working set & BFS \\
    \hline
    \hline
    \textbf{GPU Benchmarks} &  \\
    \hline
    Cache-friendly & Mini-nbody (gravitational simulation) \\
    \hline
    Cache-sensitive & Convolution, Floyd Warshall,\\
     & Recursive Gaussian \\
    \hline
    Streaming & Histogram \\
    \hline
  \end{tabular}
  \newline
  \caption{CPU-GPU benchmark classification}
  \label{benchmarks}
\end{table}
\end{scriptsize}


\section{Evaluation and Results}
In this section, we will compare CPU/GPU IPCs, LLC MPKIs, system data-bus utilization and cache area across various cache sizes and benchmark combinations for the three cache allocation policies.

\subsection{Performance}

\begin{figure}[h]
\centering
\includegraphics[width=0.48\textwidth]{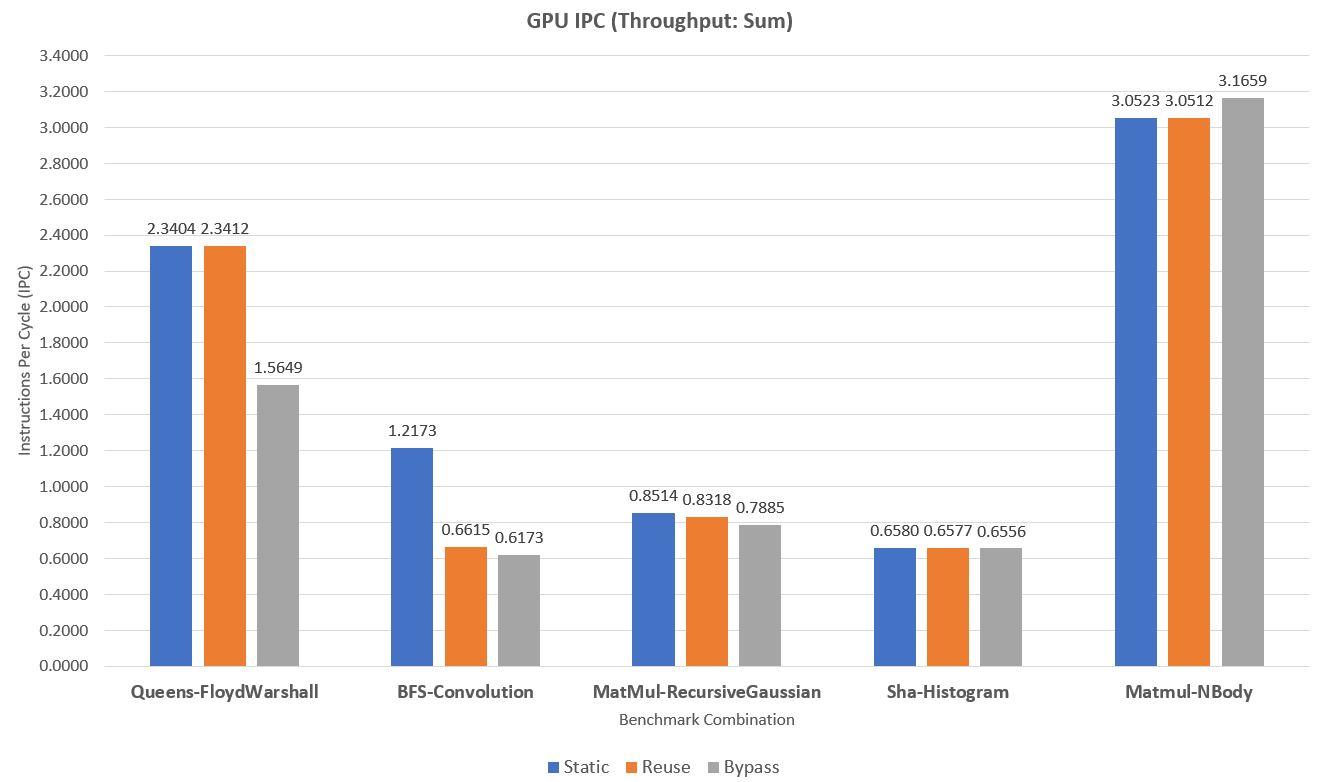}
\includegraphics[width=0.48\textwidth]{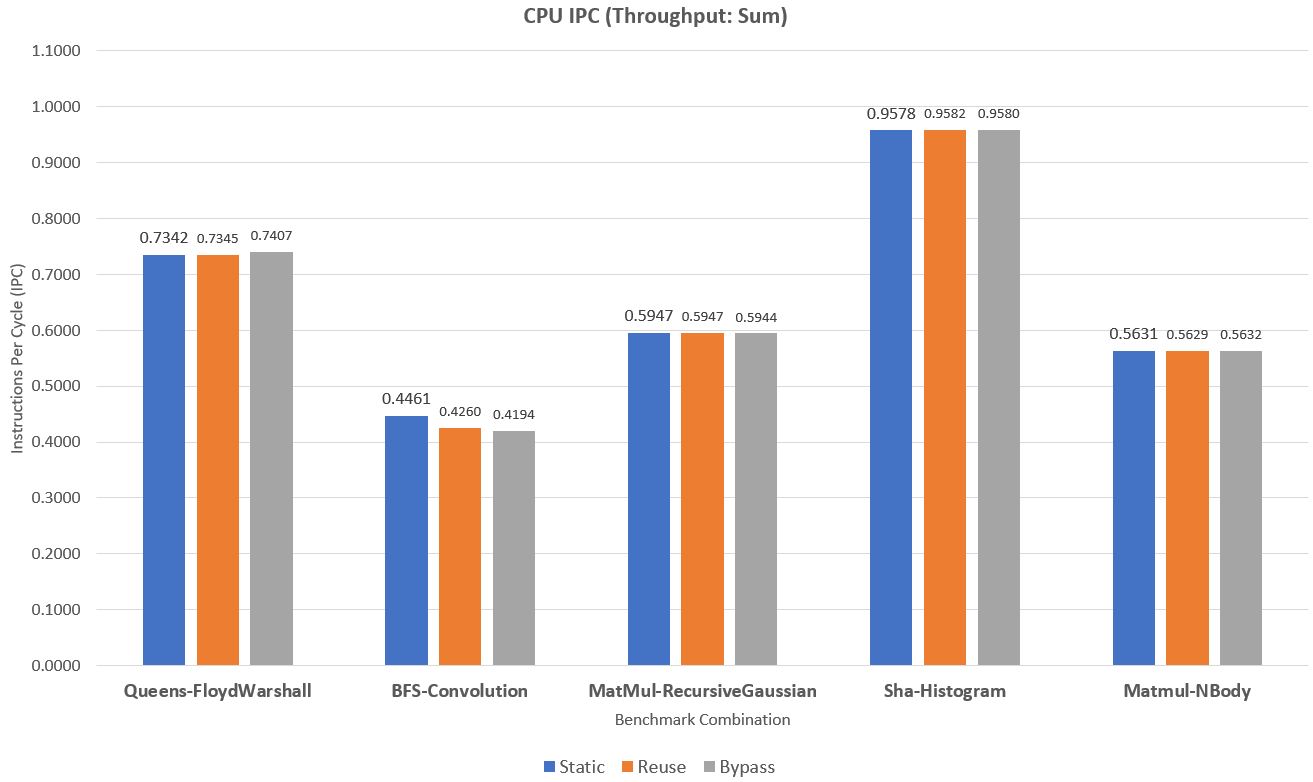}
\caption{IPC comparison across workloads}
\label{ipc}
\end{figure}

We start our evaluation by comparing the impact of these cache allocation policies on CPU and GPU performance. We use throughput as our metric, which indicates the sum of IPCs of individual cores. Figure \ref{ipc} shows the CPU-GPU IPCs across various benchmark combinations.

It can be observed that GPU IPC is always lower for LLC bypassing, except for one workload combination. On the other hand, static partitioning is significantly better (45.6\%) than the reuse cache for BFS-Convolution combination and 2\% better for Matmul-Recursive Gaussian combinations. For rest of the workloads, the reuse cache achieves mostly the same performance as static-partitioning and better compares to cache bypassing.

For the CPU IPC, bypassing slightly helps in some cases, which can be attributed to the fact that CPU access could consumes the entire LLC. On the flip side, static-partitioning and reuse cache still perform within 0.8\% of bypassing or better for all workloads.

\subsection{LLC Misses}
\begin{figure}[h]
\centering
\includegraphics[width=0.48\textwidth]{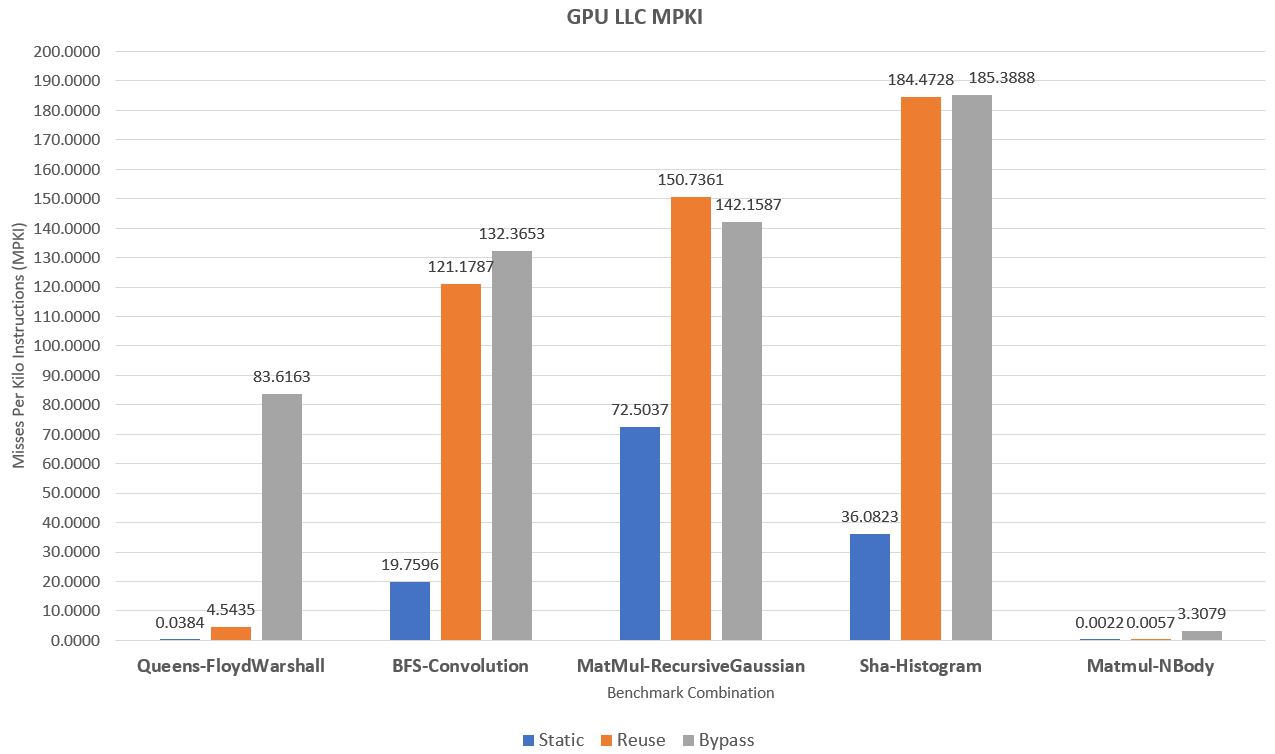}
\includegraphics[width=0.48\textwidth]{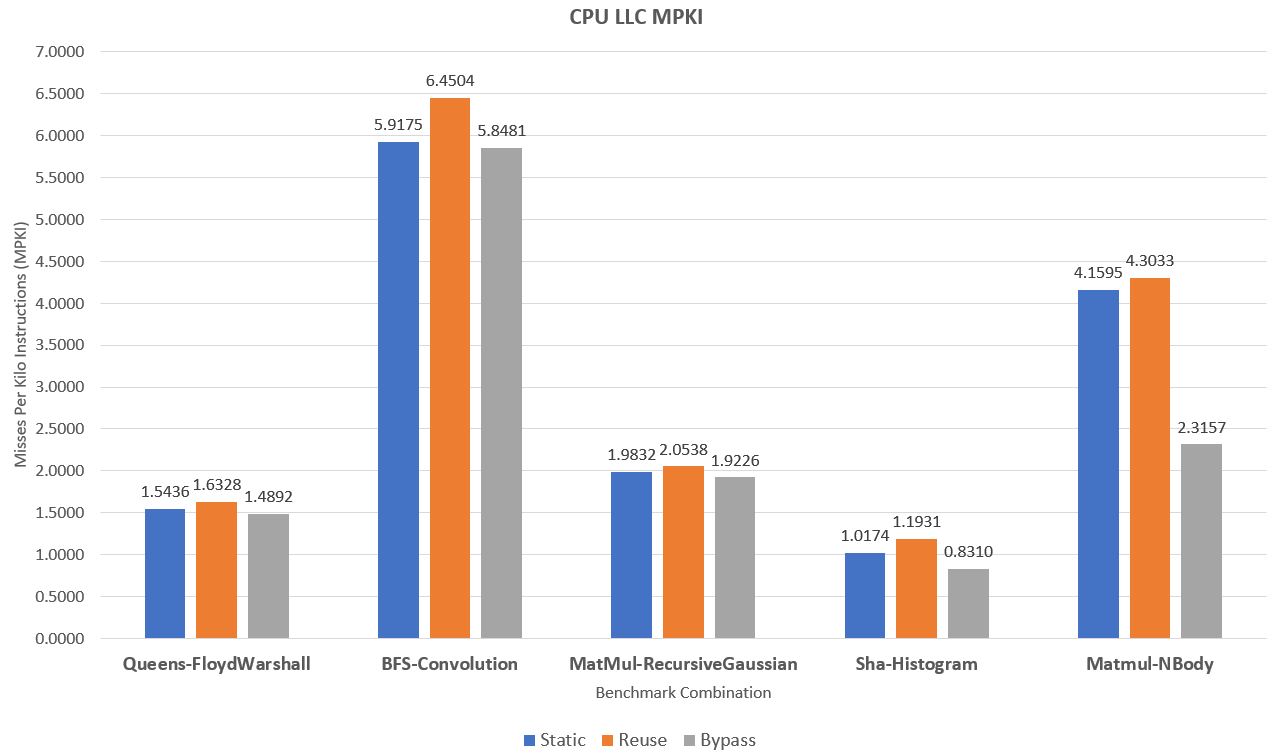}
\caption{LLC MPKI comparison across workloads}
\label{mpki}
\end{figure}

Figure \ref{mpki} shows the GPU and CPU MPKI comparison for the reuse cache and all the baselines. For CPU MPKI, both static-partitioning and the reuse cache have a higher MPKI than GPU LLC bypassing, since CPU has lesser effective cache space and thus leads to more misses. For the GPU MPKI, the reuse cache performs reasonably well, reducing GPU misses over bypassing for 3 out of 5 workloads. For the histogram workload, the reuse cache does not reduce the MPKI due to the streaming nature of the workload.

\subsection{LLC Area}
This section details the area savings provided by the reuse cache. Cacti 6.5 was used to get area numbers for conventional and reuse cache structures at 32 nm technology node.
\begin{scriptsize}
\begin{table}[h!]
  \centering
  \begin{tabular}{|l|l|}
    \hline
    \textbf{Cache Configuration} & \textbf{Area (in mm2)}\\
    \hline
    Conventional 1MB (Static/Bypass) & 2.43 \\
    \hline
    512KB Data, 1MB Tag Reuse Cache & 1.33\\
    \hline
    512KB Data, 2MB Tag Reuse Cache & 1.55\\
    \hline
  \end{tabular}
  \newline
  \caption{LLC area comparision across schemes}
   \label{area}
\end{table}
\end{scriptsize}

It is evident from table \ref{area} that the reuse cache provides 45\% and 36\% reduction in area respectively (for equal and double tag-array sizes) due to halved data-array size. The area numbers include the overhead of adding extra pointers required by tag/data entries in the reuse cache.

\subsection{Data-bus Utilization}
\begin{figure}[h]
\centering
\includegraphics[width=0.48\textwidth]{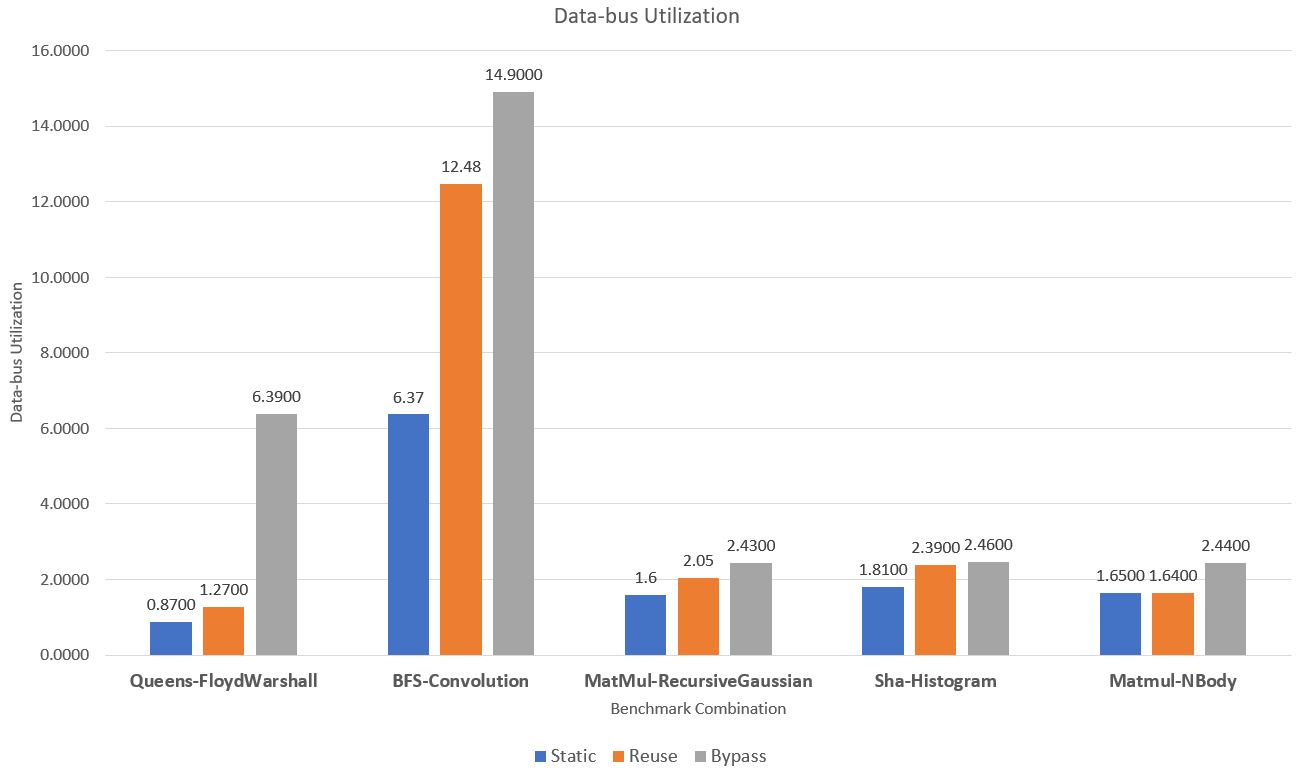}
\caption{Data-bus utilization comparison across workloads}
\label{data-bus}
\end{figure}

Figure \ref{data-bus} shows the comparison of data-bus utilization for the three cache allocation schemes. Since the GPU will bypass the LLC in cache bypassing, the DRAM accesses should also increase. This leads to GPU LLC bypassing having the maximum bus utilization out of the three cache allocation schemes. Static-partitioning has the least data-bus utilisation for most cases since it provides both CPU-GPU with a definite cache space. On the other hand, the reuse cache is able to achieve within 32\% data bus utilization for 4 of the 5 workload mixes.

\section{Conclusion}
LLC Management is an important problem in today's heterogeneous processors. The reference stream observed by the shared-LLC (SLLC) from the CPU/GPU exhibits little temporal locality, but instead, it exhibits reuse locality. As a consequence, a high proportion of the SLLC lines is useless because the lines will not be requested again before being evicted, and most hits are expected to come from a small subset of already reused lines. This proportion could be particularly high for the GPU accesses.

In this work, we looked to minimize counterproductive insertions and evictions in the SLLC by implementing a reuse cache. We evaluate our proposal by running heterogeneous CPU-GPU workloads on the gem5 AMD APU model. We use static partitioning and GPU LLC bypassing as our baseline comparison. Our major observations are: (i) Static-partitioning performs best when the GPU application working set fits into LLC and the CPU application is not cache-sensitive (ii) For the reuse cache, a performance of within 0.8\% (or better when GPU application is cache-sensitive) was achievable for most cases, providing average 40\% reduction in area (iii) A tag-to-data cache ratio of 2:1 (where data-array is half of a conventional cache) is a good starting point for reuse-cache design space exploration (iv) Simple LLC bypassing degrades performance for GPU in most cases, though it could improve CPU IPC.

Static-partitioning hard-partitions the cache, while GPU LLC-bypassing does not provide space to GPU in LLC. We think reuse cache incorporates properties from both, hence could adapt to a wider range of applications.

\section{Future Work}
In this paper, we have achieved comparable or slightly better application performance with reuse cache in most cases, at significant area reduction. Further reuse cache performance improvements might be closely linked with data/tag replacement policies and memory-access rate (and thread-awareness) in heterogeneous systems. Potential performance improvement might also be obtained by adding a “Tag-only” coherence state in the reuse cache coherence protocol, and increasing the reuse hysteresis. Reuse cache could be combined with cache compression techniques where we have more tags than data blocks, thereby reducing data-array size further.


\bibliographystyle{IEEEtranS}


\section*{Appendix}
\begin{scriptsize}
\begin{table}[h!]
  \centering
  \begin{tabular}{|l|l|}
    \hline
    \textbf{Name} & \textbf{Contribution}\\
    \hline
    \hline
    Tejas Shah & Static-partitioning implementation, \\
    & Evaluation and benchmark analysis \\
    \hline
    Bobbi Yogatama & Reuse cache implementation, \\
    & GPU LLC bypassing implementation \\
    \hline
    Kyle Roarty & gem5 setup for running CPU-GPU programs, \\ 
    & Benchmark setup\\
    \hline
    Rami Dahman & Benchmark setup, \\ 
    & Evaluation and benchmark analysis\\
    \hline
  \end{tabular}
  \newline
  \caption{Individual Contributions of Team Members}
\end{table}
\end{scriptsize}

\end{document}